\documentclass[apj]{emulateapj}

\slugcomment{{\it Submitted 7 November 2005; accepted 27 February 2006}}

\shorttitle{Rejecting Astrophysical False Positives from TrES}
\shortauthors{O'Donovan et al.}

\newcommand{\gsc}{\mbox{GSC\,03885--00829}}

\newcommand{\first}{\mbox{HD\,209458\,b}}

\begin{document}

\title{Rejecting Astrophysical False Positives from the \\ T\lowercase{r}ES
  Transiting Planet Survey: \\ The Example of \gsc \altaffilmark{1} }

\author{Francis~T.~O'Donovan\altaffilmark{2},
  David~Charbonneau\altaffilmark{3}, Guillermo~Torres\altaffilmark{3},
  Georgi~Mandushev\altaffilmark{4}, Edward~W.~Dunham\altaffilmark{4},
  David~W.~Latham\altaffilmark{3}, Roi~Alonso\altaffilmark{5},
  Timothy~M.~Brown\altaffilmark{6},
  Gilbert~A.~Esquerdo\altaffilmark{7},
  Mark~E.~Everett\altaffilmark{7}, Orlagh~L.~Creevey\altaffilmark{6}}

\altaffiltext{1}{Some of the data presented herein were obtained at
  the W.M. Keck Observatory, which is operated as a scientific
  partnership among the California Institute of Technology, the
  University of California, and the National Aeronautics and Space
  Administration. The Observatory was made possible by the generous
  financial support of the W.M. Keck Foundation.}

\altaffiltext{2}{California Institute of Technology, 1200 East
  California Boulevard, Pasadena, CA 91125; ftod@caltech.edu}

\altaffiltext{3}{Harvard--Smithsonian Center for Astrophysics, 60
  Garden Street, Cambridge, MA 02138}

\altaffiltext{4}{Lowell Observatory, 1400 West Mars Hill Road,
  Flagstaff, AZ 86001}

\altaffiltext{5}{Instituto de Astrof\'{\i}sica de Canarias, 38200 La
  Laguna, Tenerife, Spain}

\altaffiltext{6}{High Altitude Observatory/National Center for
  Atmospheric Research, 3080 Center Green, Boulder, CO 80301}

\altaffiltext{7}{Planetary Science Institute, 1700 East Fort Lowell
  Road, Suite 106, Tucson, AZ 85719}

\begin{abstract}
  Ground--based wide--field surveys for nearby transiting gas giants
  are yielding far fewer true planets than astrophysical false
  positives, of which some are difficult to reject. Recent experience
  has highlighted the need for careful analysis to eliminate
  astronomical systems where light from a faint eclipsing binary is
  blended with that from a bright star. During the course of the
  Trans--atlantic Exoplanet Survey, we identified a system presenting
  a transit--like periodic signal. We obtained the proper motion and
  infrared color of this target (\gsc) from publicly available
  catalogs, which suggested this star is an F dwarf, supporting our
  transit hypothesis.  This spectral classification was confirmed
  using spectroscopic observations from which we determined the
  stellar radial velocity. The star did not exhibit any signs of a
  stellar mass companion. However, subsequent multi--color photometry
  displayed a color--dependent transit depth, indicating that a blend
  was the likely source of the eclipse. We successfully modeled our
  initial photometric observations of \gsc\ as the light from a K
  dwarf binary system superimposed on the light from a late F dwarf
  star. High--dispersion spectroscopy confirmed the presence of light
  from a cool stellar photosphere in the spectrum of this system. With
  this candidate, we demonstrate both the difficulty in identifying
  certain types of false positives in a list of candidate transiting
  planets and our procedure for rejecting these imposters, which may
  be useful to other groups performing wide--field transit surveys.
\end{abstract}


\section{Introduction}
\label{sec:intro}

The discovery of the $169$ known extrasolar planets%
\footnote{Updates available from the 
\anchor{http://www.obspm.fr/planets/}{Extrasolar Planets Encyclopaedia}:\\
\url{http://www.obspm.fr/planets/}.}%
\ has greatly enhanced our
understanding of planetary systems. Most of these extrasolar planets
have been identified from Doppler surveys, which search for the radial
velocity variation of a star caused by the presence of a gas giant.
These observations can estimate the period and eccentricity of the
planetary orbit, and provide a lower limit for the planetary mass
relative to that of the star. Assuming that the stellar mass can be
precisely determined, perhaps by comparing the stellar spectra to
stellar atmosphere models (\citealp[such as][]{Kurucz:ATLAS9:1993a}),
the minimum mass of the planet can be derived.  The diversity of these
planetary systems has drastically altered our theoretical appreciation
of their morphology and evolution, including the existence of ``hot
Jupiters'', Jupiter--sized planets with periods of a few days that
experience high insolation from the nearby star.  However, less is
known about the nature of the planets themselves, except for the cases
when a hot Jupiter was observed to pass in front of a dwarf star. By
observing such a transit (as was first suggested by
\citealt{Struve:obs:1952a}), we can estimate the radius of the
transiting planet from the fraction of starlight the planet blocks
during the transit and the radius of the star itself, which must be
measured from stellar model fits to spectra of the star. For a transit
to occur, the planetary orbital inclination must be approximately
$90\arcdeg$; this implies that the Doppler limit for the planetary
mass must be close to the actual mass of the planet.  We now have
estimates for the radii and masses of 9 planets from transit
observations similar to those first successfully performed by
\cite{Charbonneau_Brown_Latham:apjl:2000a} and
\cite{Henry_Marcy_Butler:apj:2000a}. These estimates provide
constraints for the competing theories of planetary formation. For
example, the recently discovered transiting ``hot Saturn''
\mbox{HD\,149026\,b} has a mass and radius that imply the
planet has a large core (approximately $70\,\mathrm{M}_{\earth}$;
\citealt{Sato_Fischer_Henry:apj:2005a,Charbonneau_Winn_Latham:apj:2006a}).
It is hypothesized that this planet must have formed via core
accretion \citep{Pollack:araa:1984a,Pollack_Hubickyj_Bodenheimer:icarus:1996a},
rather than through gravitational instability \citep{Boss:sc:1997a}.
However, not every observation of a transiting planet can be explained
by current models of close--in giant planets experiencing high stellar
insolation. The transiting planet \first\ has a radius larger than
other planets of its mass, and larger than the predicted radii
(\citealp[see][ and references
therein]{Laughlin_Wolf_Vanmunster:apj:2005a}, and
\citealt{Deming_Seager_Richardson:nat:2005a}).

Several wide--field photometric surveys with the goal of identifying
nearby transiting planets are currently active. Our
\anchor{http://www.astro.caltech.edu/~ftod/tres/}{Trans--atlantic
  Exoplanet Survey}%
\footnote{\url{http://www.astro.caltech.edu/\textasciitilde ftod/tres/}}%
\ (TrES) is a network of three 10\,cm telescopes:
\anchor{http://www.astro.caltech.edu/~ftod/tres/sleuth.html}{Sleuth}
 (located at Palomar Observatory, California), PSST (Lowell
Observatory, Arizona; \citealt{Dunham_Mandushev_Taylor:pasp:2004a}),
and
\anchor{http://www.hao.ucar.edu/public/research/stare/stare.html}{STARE}%
\footnote{\url{http://www.hao.ucar.edu/public/research/stare/stare.html}}%
\ (Tenerife, Spain; \citealt{Alonso_Deeg_Brown:an:2004a}). The TrES
campaign, together with other wide--field surveys such as the HAT
network \citep{Bakos_Lazar_Papp:pasp:2002a} and SuperWASP
\citep{Street_Pollaco_Fitzsimmons:ASP:2003a}, are monitoring thousands
of nearby bright stars ($9\leq V \leq 13$). We hope to find recurring
eclipses with the short period and small amplitude corresponding to a
transiting hot Jupiter. The brightness of the target stars facilitates
both the photometric precision and the follow--up of any identified
transiting planets using space--borne telescopes. Examples of detailed
follow--up observations include the measurement of several chemical
abundances in the atmosphere of \first\
\citep{Charbonneau_Brown_Noyes:apj:2002a,Vidal-Madjar_Lecavelier-des-Etangs_Desert:nat:2003a}
and the first direct detections of emitted planetary radiation
\citep{Charbonneau_Allen_Megeath:apj:2005a,Deming_Seager_Richardson:nat:2005a}.

Many astrophysical systems exist that mimic the light curve of a
transiting hot Jupiter. Due to the mass--radius degeneracy for bodies
with masses between 0.001 and 0.1\,M$_{\sun}$, the depth of an eclipse
of a solar--type star by a hot Jupiter, a brown dwarf or an M dwarf
star will be about 1\% in each case, despite the large range in mass.
Grazing incidence eclipsing binaries may also exhibit comparable
eclipse depths. For wide--field ground--based surveys, the frequency
of the false positives is greater than the frequency of detection of
true transiting planets, by at least an order of magnitude.  As part
of TrES, we typically identify 10--20 of these transit--like
photometric signals out of 15,000--25,000 stars ($10<V<15$) in each
$6\arcdeg \times 6\arcdeg$ target field of view ($b\sim15\arcdeg$) (%
\citealp[see, e.g.,][]{Dunham_Mandushev_Taylor:pasp:2004a}), and
similar yields should be expected from other wide--field ground--based
surveys. This number (which is dependent on the density of star
counts, and hence the Galactic latitude) is consistent with
theoretical predictions. For example, \cite{Brown:apjl:2003a} predicts
that for every 25,000 stars observed, we will find 10 false positives
and only one true transiting planet. This assumes that we must observe
3 transit events for each candidate, and is dependent on the
visibility of transits throughout the observation run. The low yield
of planets necessitates a rigorous routine of follow--up observations
and detailed analysis to eliminate all possible alternatives to the
planet hypothesis. One straightforward method to reduce the number of
false positives is to obtain multi--epoch spectroscopy of each
candidate and measure radial velocities with a precision of $\sim
1\,\mathrm{km\,s^{-1}}$. From this we can identify targets with
companions of stellar, rather than planetary, mass (%
\citealp[see,
e.g.,][]{Latham:ASP:2003a,Charbonneau_Brown_Dunham:AIP:2004a}). We can
also estimate the luminosity class of the target star to single out
and reject giants.

The blending of the light from an edge--on binary system with that
from a third star can also be mistaken for a transit.
\cite{Brown:apjl:2003a} predicts that half of the false positives from
a typical wide--field survey will be of this type; the other half will
be grazing eclipsing binaries. Blends can be much more difficult to
identify. The faintness of the binary compared to the third star can
prevent the detection of the radial velocity variations of the
binary. These variations are also masked by the rotationally broadened
spectral lines of the rotationally synchronized binary stars. However,
if the binary has a significant difference in effective temperature
from that of the third star, the eclipse depths should display a
strong color dependence, unlike the color--independent transits of a
(dark) planet across a single star.  Nevertheless, there has been
recent experience of blends with color--independent eclipse depths. In
the case of \mbox{OGLE--TR--33}
\citep{Torres_Konacki_Sasselov:apj:2004b} and \mbox{GSC\,01944--02289}
\citep{Mandushev_Torres_Latham:apj:2005a}, both candidates (a
suspected planet and brown dwarf, respectively) showed
color--independent eclipse depths, and yet were subsequently
discovered to be blended systems. Evidence for the presence of an
eclipsing binary was found from a careful analysis of the spectral
line shapes, prompting the authors to compare the photometric data to
simulations of blends. \mbox{OGLE--TR--33} was shown to
be a hierarchical triple consisting of a bright F6 dwarf and an
F4+(K7--M0) binary. The blend model for \mbox{GSC\,01944--02289} comprises an F5 primary and a G0+M3
binary. In both cases, the similarity in color between the primary
star and the brightest member of the binary explains the constant
eclipse depth at different wavelengths. The high occurrence of such
false positives and the difficulty in rejecting them requires a
detailed study of candidates before any announcement is made, as was
done in the case of \mbox{TrES--1}
\citep{Alonso_Brown_Torres:apjl:2004a} and \mbox{OGLE--TR--56\,b} \citep{Torres_Konacki_Sasselov:apj:2004a}.

Here we discuss a promising candidate, \gsc, from one of our target
fields. Initial photometric (\S\ref{sec:tres}) and spectroscopic
(\S\ref{sec:spec}) monitoring of this candidate strongly suggested
that we were observing a Saturn--sized companion transiting a solar
type star every 1.441 days with a transit depth of approximately
6\,mmag. However, follow--up photometry (\S\ref{sec:photo}) displayed
a slight color dependence as might be caused by a blend, and we were
able to model our photometry using simulations of blended eclipsing
binaries (\S\ref{sec:blend}). Our best fit model consists of a bright
F dwarf and a K dwarf binary, and we were able to identify the presence 
of light from the binary in the spectrum of \gsc\ (\S\ref{sec:nirspec}). 
The faintness of the binary system 
prevents us from identifying the presence of asymmetric spectral
lines, as was done by \cite{Torres_Konacki_Sasselov:apj:2004b} and
\cite{Mandushev_Torres_Latham:apj:2005a}. In this case, only
multi--color observations provided us with the necessary evidence to
call into question the planetary nature of this candidate. 

\section{TrES Telescope Observations}
\label{sec:tres}

In March 2004, we commenced observations of a $6\arcdeg\times6\arcdeg$
target field in Draco. The field is centered on our $V=4.8$ guide star
\mbox{HD\,151613} ($\alpha = 16^{\rm h} 45^{\rm m}
17\fs 82$, $\delta = +56\arcdeg 46\arcmin 54\farcs7$ J2000). Between
UT 2004 March 29 and June 22, we observed this field nightly with
\anchor{http://www.astro.caltech.edu/~ftod/tres/sleuth.html}{Sleuth}
at Palomar Observatory (California), and with PSST at Lowell
Observatory (Arizona). STARE, in Tenerife (Spain), did not observe
this field as it was undergoing repairs at this time. A total of 15854
photometric exposures of 90\,s each were obtained through either a
Sloan $r$ filter (Sleuth) or a Kron-Cousins $R$ filter (PSST).

We bias--subtracted and flat--fielded the images of our target field
once the data were transferred from the observatory computers. We
performed the calibration of the Sleuth data using customized IDL
routines; we calibrated the PSST data using the \texttt{zerocombine},
\texttt{ccdproc}, and \texttt{flatcombine} tasks in the IRAF%
\footnote{IRAF is distributed by the National Optical Astronomy
  Observatories, which are operated by the Association of Universities
  for Research in Astronomy, Inc., under cooperative agreement with
  the National Science Foundation.}%
\ package \citep{Tody:1993a}. We reduced the Sleuth and PSST
photometric data separately as follows using our difference image
analysis (DIA) pipeline (%
\citealp[described in][]{Dunham_Mandushev_Taylor:pasp:2004a}, 
\citealp[and based in part upon][]{Alard:aas:2000a}).

We created our reference image for the field from images obtained at
low air mass on a photometric night during dark time. We obtained a
standard list of stars from this image using profile--fitting (PSF)
photometry (DAOPHOT II/ALLSTAR;
\citealt{Stetson:pasp:1987a,Stetson:ASP:1992a}). We calculated the
equatorial coordinates ($\alpha,\delta$) of these stars by matching a
subset with the stars listed in the Tycho--2 Catalog
\citep{Hog_Fabricius_Makarov:aa:2000a}, and then spatially
interpolated all of the science images so that the star coordinates
from each image matched those from our standard star list.

We produced the master image for the Sleuth data set by combining 19
of our best--quality interpolated images; we combined 17 images to
create the corresponding PSST master image. We subtracted each
interpolated image from this master image. We used aperture photometry
on the resultant difference images (using the centroids derived for
the standard star list) to estimate the flux of each star in each
image. We produced time series consisting of the differences between
the magnitude of a star in the reference image and the magnitude of
that star in each target image in turn. We decorrelated these light
curves as follows to remove systematic effects typical of wide--field
surveys, such as those caused by changing atmospheric conditions
throughout the night. We listed the stars in order of brightness, and
divided the list into batches of 500. We computed the least--squares
fit to the light curve of a given star from a linear combination of
the other light curves in that batch. We then subtracted this
least-squares fit from the light curve of the star.

In our previous studies
\citep{Alonso_Brown_Torres:apjl:2004a,Mandushev_Torres_Latham:apj:2005a,Creevey_Benedict_Brown:apjl:2005a},
we presented separate light curves from one of the TrES telescopes.
For this field, we combined the two data sets. For a given star on the
Sleuth standard star list, we calculated the angular distances between
that star and the PSST standard stars. We matched the Sleuth star with
a PSST star if the angular distance was less than $5\arcsec$ (0.5
pixels). Due to the difference in the selected filter and the
telescope pointing between the two sites, some stars were unique to a
given standard star list, and no match was found. We appended the time
series for each matched PSST star to the corresponding Sleuth time
series, and reordered the combined time series chronologically. The
data for the unmatched PSST stars were simply added to the resultant
data set.

In order to reduce the computational intensity of our transit search,
we averaged the combined time series in 9 minute wide bins to obtain
2996 binned observations. Since central transits should last 3 hours,
this did not significantly sacrifice temporal resolution of potential
transit events. For $\sim$10,000 stars, the rms scatter of the binned
data was below 0.04 mag . We performed a search of the time series of
these stars using the box--fitting least squares transit--search
algorithm (BLS; \citealt*{Kovacs_Zucker_Mazeh:aa:2002a}) to identify
periodic transit events. The BLS algorithm calculates a Signal
Detection Efficiency (SDE;
\citealp[see][]{Kovacs_Zucker_Mazeh:aa:2002a}) for each candidate,
which denotes how significant the detection is. We identified
candidates based on this SDE, followed by a visual inspection.

\begin{figure}[p]
\epsscale{1.0}
\plotone{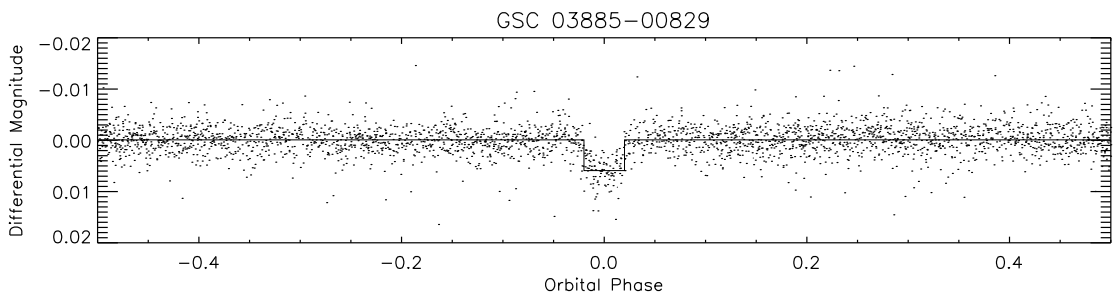}
\caption{Binned TrES $r$ light curve of \mbox{GSC\,03885--00829},
  folded with the photometric period of 1.441 days computed using the
  Box--fitting Least Squares algorithm. Overlaid is the corresponding
  box transit model.}
\label{fig:discovery}
\end{figure}

Many of the candidates identified for this field show V-shaped
eclipses or a possible ellipsoidal variability, or have large depths,
making it likely that they are not transiting planets, but rather
eclipsing binaries.  However, we promptly identified a promising
candidate. When the data for this star were folded with the
photometric orbital period of 1.441 days (calculated using the BLS
algorithm), the resultant light curve (Figure~\ref{fig:discovery})
displayed a shallow and flat--bottomed transit, and no noticeable
ellipsoidal variability out of transit. The depth ($\sim$6\,mmag) and
duration (1.4 hours) of the occultation are consistent with a
Saturn--sized planet transiting a solar type star.  The SDE for this
candidate ($\sim 20$) was high relative to that calculated for typical
TrES candidates ($\sim10$--15).

Supporting data from online catalogs provided further evidence of the
planetary nature of this eclipse. Using the
\anchor{http://simbad.harvard.edu/}{SIMBAD}%
\footnote{\url{http://simbad.harvard.edu/}}%
\ database, we identified our candidate as the star \gsc\ (see
Table~\ref{tab:gsc}). The infrared colors (2MASS $J-K$, $J-H$;
\citealt{Cutri_Skrutskie_van-Dyk:VizieR:2003a}) and optical ($B-V$)
colors of this star are near--solar, roughly consistent with the
stellar parameters inferred from our transit observations. This star
displays significant proper motion (26 mas/year from the USNO--B
Catalog; \citealt{Monet_Levine_Canzian:aj:2003a}), suggesting it is a
nearby dwarf. As a first check of the possibility of contamination of
light from a nearby star, we verified that there is no bright star
visible on the Digitized Sky Survey%
\footnote{The \anchor{http://archive.stsci.edu/dss/}{Digitized Sky
    Survey} (\url{http://archive.stsci.edu/dss/}) was produced at the
  Space Telescope Science Institute under U.S. Government grant NAG
  W-2166. The images of these surveys are based on photographic data
  obtained using the Oschin Schmidt Telescope on Palomar Mountain and
  the UK Schmidt Telescope. The plates were processed into the present
  compressed digital form with the permission of these institutions.}%
\ (DSS) images within our aperture radius ($\leq 30\arcsec$).

With due enthusiasm, we proceeded to obtain follow--up observations of
this exciting candidate, with the goal of rejecting the possibility
that this was not a transiting planet.

\section{Spectroscopic Follow--up}
\label{sec:spec}

We confirmed that \gsc\ was an isolated dwarf star by
spectroscopically monitoring this candidate. We observed \gsc,
together with other candidates from this TrES field, with the
Harvard--Smithsonian Center for Astrophysics (CfA) Digital Speedometer
\citep{Latham:ASP:1992a}, operated on the 1.5\,m Tillinghast reflector
at the F.\ L.\ Whipple Observatory (FLWO) on Mt.\ Hopkins, Arizona.
The spectral coverage was 45\,\AA\ centered on 5187\,\AA\ at a
resolving power of $\lambda / \Delta \lambda \approx 35,\!000$ (a
resolution of $8.5\,\mathrm{km\,s^{-1}}$). We observed this particular
target on UT 2005 May 18, May 20 and May 21, at an orbital phase 0.52,
0.86, and 0.50, respectively (calculated using the orbital ephemeris
of the planet; see Table~\ref{tab:gsc}c).

\begin{figure}[p]
\epsscale{0.65}
\plotone{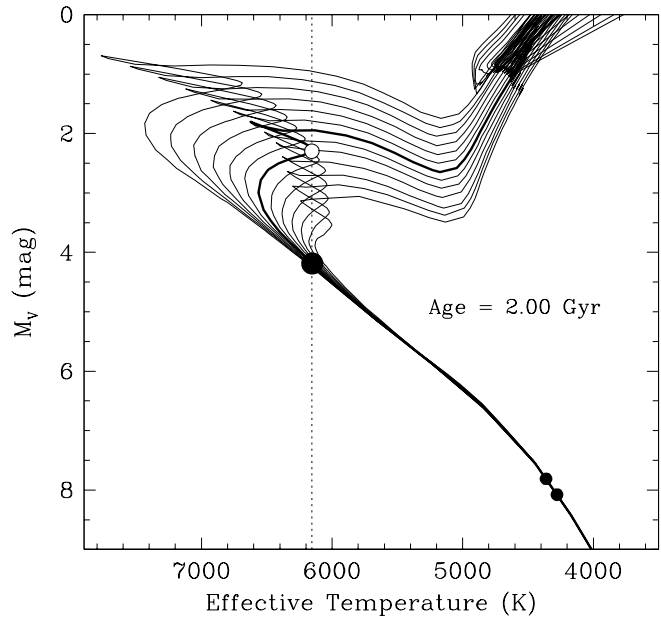}
\caption{\cite{Girardi_Bressan_Bertelli:aas:2000a} model isochrones
  for solar metallicity and ages ranging from 1 to 4 Gyr. The open
  circle and large filled circle represent two main-sequence stars of
  the same effective temperature as our candidate
  \mbox{GSC\,03885--00829} ($T_{\rm eff} = 6150$ K), but different
  degrees of evolution. They are shown on the 2 Gyr isochrone (heavy
  line), which maximizes the difference in brightness at this
  temperature. The corresponding radii differ by a factor of 2.4. We
  must therefore constrain the evolution of our candidate using
  spectroscopy before we can accurately estimate its radius and hence
  the size of any transiting companion. The three filled circles show
  the location of each of the three members of our final blend model
  for this candidate on the 2\,Gyr isochrone. The bright primary has a
  mass of 1.15\,M$_{\sun}$, and the binary component masses are
  0.67\,M$_{\sun}$ and 0.64\,M$_{\sun}$.}
\label{fig:isochrones}
\end{figure}

Radial velocities were obtained by cross-correlation using templates
chosen from a library of synthetic spectra computed for us by J.~Morse
and based on the model atmospheres of R.~L.~Kurucz (J.~Morse \&
R.~L.~Kurucz, 2004, private communication). The typical precision of a
single velocity measurement is $0.5\,\mathrm{km\,s^{-1}}$. We measured
the radial velocity to be constant ($-38.48\,\mathrm{km\,s^{-1}}$ with
an rms of $0.28\,\mathrm{km\,s^{-1}}$) within our errors. These
measurements indicate that the target star is not gravitationally
bound to a massive stellar companion. Various stellar parameters were
estimated, again by cross-correlating these spectra against a grid of
templates from our spectral library, seeking the best match.  Assuming
a solar metallicity, we estimated the effective temperature to be
$T_{\mathrm{eff}}=6150$\,K and the surface gravity to be $\log{g}
\approx 4.4$, which suggested this was a late F dwarf star, consistent
with the proper motion and photometric colors, and with our transit
hypothesis. The formal stellar rotation we derived ($v\sin{i} \approx
1\,\mathrm{km\,s^{-1}}$) is actually below our spectral
resolution. The surface gravity suggests the star is unevolved. This
constraint is important, as this star lies within the range of
effective temperatures for which an ambiguity exists as to the
corresponding mass of the star while on the main sequence, depending
on the degree of evolution. An illustration of this ambiguity is shown
in Figure~\ref{fig:isochrones}, where the two locations denoted by the
open circle and large filled circle have the same effective
temperature as \gsc\ but rather different luminosities. The particular
age of the isochrone for this figure was selected to show this
difference more clearly. The fainter (lower) location corresponds to
an unevolved main--sequence star ($\log{g}=4.34$) of mass
1.15\,M$_{\sun}$, whereas the brighter location is for a star near the
end of the hydrogen--burning phase, and has a surface gravity of
$\log{g}=3.68$ and a mass of 1.61\,M$_{\sun}$. The radii differ by a
factor of about 2.4. From this example, we see that, without a surface
gravity constraint, we cannot be sure of the radius of our target
star, preventing our discriminating between a planetary and a stellar
transiting companion.

\section{Photometric Follow--up}
\label{sec:photo}

Although we had found no evidence of a stellar mass companion from our
radial velocity measurements, the possibility remained that the
observed transits were in fact the eclipses of a faint binary whose
light was blended with that from the bright F dwarf due to the large
pixel sizes of our detectors. Follow--up observations of higher
angular resolution might resolve such a blended system. A possible
wavelength dependence of the eclipse depth from multi--color
observations would also provide evidence of a blend. We organized a
follow--up photometric campaign to observe multiple transits of \gsc\
using $D>20\arcsec$ (0.50\,m) telescopes and through several different
filters.

\begin{figure}[p]
\begin{center}
\includegraphics[scale=0.5]{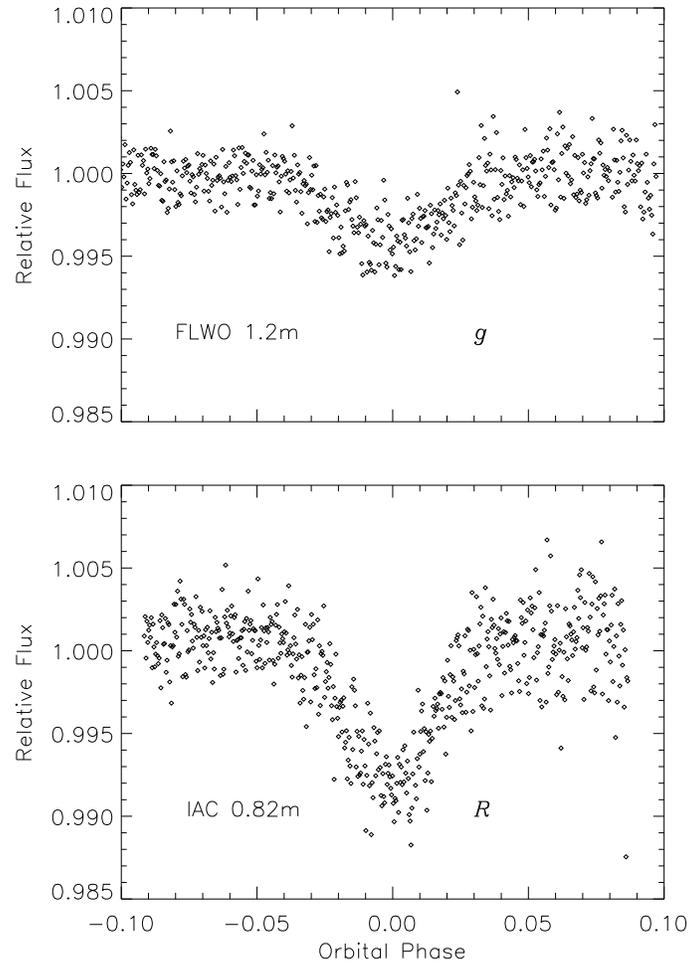}
\end{center}
\caption{Our follow--up photometry of \mbox{GSC\,03885--00829} near
  the predicted time of transit. The observations were folded using
  the orbital ephemeris of the planet (see Table~\ref{tab:gsc}c). Each
  plot is labelled with the corresponding telescope (see text), and
  the filter bandpass. The predicted transit events were observed, but
  an increase in eclipse depth with increasing wavelength is
  apparent.}
\label{fig:multicolor}
\end{figure}

High precision photometry of \gsc\ was made on UT 2005 June 8 using
the 1.2\,m FLWO telescope (Arizona).  For this clear, photometric
night, we used MiniCam, a two CCD mosaic, each array being $2048
\times 4608$ pixels.  Observations were binned $2\times2$ for a faster
duty cycle. A Sloan $g$ filter was used, with an exposure time of
30\,s and a corresponding duty cyle of $\sim$50.5\,s.  The telescope
was de--focused to a FWHM of $\sim$15 pixels ($\sim$9$\arcsec$) to
allow greater photon counts per given exposure without saturation.
Spreading the star image also serves to reduce pixel--to--pixel
variations that may not be completely removed by flat--fielding. A
total of 561 photometric measurements of the field were made over a
total of 7.875 hours. The differential light curve of our target star
was obtained using aperture photometry of this star and one of the
observed reference stars. We corrected for the effect of differential
extinction on the time series by fitting the out--of--eclipse
data. The resultant light curve was then converted to flux units (see
Figure~\ref{fig:multicolor}). These follow--up photometric
observations were made over a year after our TrES images. We used this
separation in time to obtain a more accurate photometric ephemeris for
our candidate, $T_{c} (\mathrm{HJD}) = 2453529.833 + 1.44122 \times E$
(see Table~\ref{tab:gsc}).

\gsc\ was observed on the night of UT 2005 June 13 at the 0.82\,m
IAC--80 telescope at the Observatorio del Teide (Tenerife, Spain),
using its 1024$\times$1024 pixels CCD and a Johnson $R$ filter. To
achieve better photometric precision, a slight defocus was applied, so
as to image stars with a FWHM of $\sim$4$\arcsec$.  Exposures times of
22\,s were used, and the readout time using a 2$\times$2 binning was
$\sim$10\,s. The images were bias-- and flat--field corrected, and
aperture photometry was applied using the package for optimal aperture
photometry \texttt{vaphot} \citep{Deeg_Doyle:2001a}. Nine reference
stars were used to build an essemble reference star. The dispersion of
the data points is larger at the end of the night as the star was
closer to the horizon. We corrected for differential extinction;
Figure~\ref{fig:multicolor} shows the derived differential light
curve.

We obtained $BV(RI)_{\rm C}$ observations of \gsc\ on UT 2005 June 5
with the Lowell Observatory $42\arcsec$ (1.05\,m) Hall reflector in
combination with a $2{\rm K} \times 2{\rm K}$ SITe CCD. A total of 223
exposures of the program field were accumulated: four in $B$, 211 in
$V$, three in $R_C$, and five in $I_C$. We observed 20 photometric
standards in the SA107, SA108, SA109, SA112, and PG1633 fields
\citep{Landolt:aj:1992a} in order to calibrate the photometry. We
obtained the following values%
\footnote{The errors include the uncertainties in the Landolt
  photometry and the internal scatter of our photometry, but may not
  account for the total systematic error in the observations.}%
\ for the standard magnitudes of \gsc\ (the numbers in the brackets
show the number of individual measurements used to derive the mean
magnitudes): $B = 11.077 \pm 0.001\ (4)$, $V = 10.465 \pm 0.001\
(12)$, $R_{\rm C} = 10.094 \pm 0.003\ (3)$, and $I_{\rm C} = 9.722 \pm
0.001\ (5)$.

These initial attempts to observe \gsc\ convinced us that the target
star displayed a transit--like dip and that nearby stars were not
variable. This reduced the possibility that this signal was caused by
a chance superposition of a star with an eclipsing binary in the large
pixel scale of the TrES detectors. We were also able to reproduce our
$g$ band observations using a model of a Jupiter--sized planet
orbiting a near--solar type star, although the ingress and egress
appeared to be too long in duration.

It was when we compared the different light curves that we realized
something was wrong with our assumption that these were observations
of a transiting planet. The depth of the transit appeared to vary with
wavelength: an eclipse depth of 0.4\% in the $g$ band and 0.7\% in the
$R$ band. Prompted by the color dependence of the transit depths, we
ran various simulations of the eclipses visible from \gsc\ in an
attempt to rule out the possibility of a blend.

\section{Blend Analysis}
\label{sec:blend}

\begin{figure}[p]
\epsscale{1.0}
\plotone{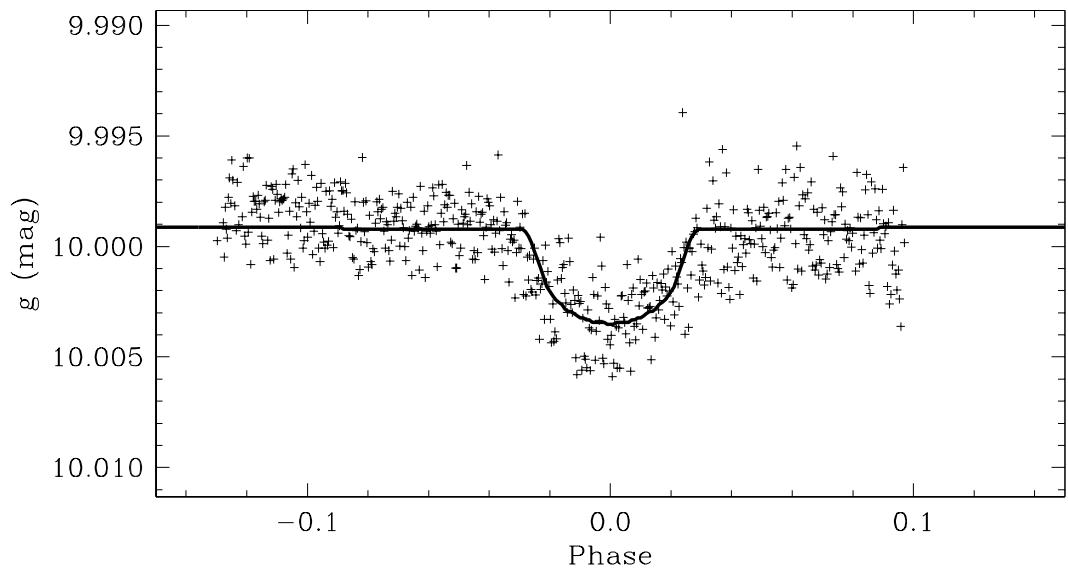}
\caption{Binned $g$-band observations of \mbox{GSC\,03885--00829}
  obtained with the FLWO 1.2\,m telescope, folded with the BLS period
  of 1.441 days. Superimposed is the theoretical light curve from a
  blend model consisting of the bright F star and a K+M dwarf
  eclipsing binary (see text). Although the depth of the transit is
  well fit, the predicted duration is slightly shorter than observed.}
\label{fig:KMmodel}
\end{figure}

Light curve fits to the $g$--band observations were carried out as
described in detail by
\cite{Torres_Konacki_Sasselov:apj:2004b}. Briefly, we assumed that the
measured brightness of \gsc\ is due to the light of an eclipsing
binary blended with the light of the F star, so that the deep eclipses
of the binary are reduced in depth to the level that we see. We
hypothesized that the three objects formed a hierarchical triple
system (rather than a by--chance alignment), and we took their
physical properties from theoretical isochrones by
\cite{Girardi_Bressan_Bertelli:aas:2000a}. The mass of the F star
(1.15\,M$_{\sun}$) was constrained from its effective temperature
derived in \S\ref{sec:spec}, with the assumption that it is unevolved.
This modeling produced a reasonably good match to the measured dip for
an edge--on binary composed of an early K star eclipsed by a very
small M star (see Figure~\ref{fig:KMmodel}).  In this scenario there
is no measurable secondary eclipse.  Although the eclipse depth is
well reproduced, the predicted duration is slightly shorter than that
observed.  In addition, the brightness expected for the K star
($\sim$15\% of the F star in the optical) is such that it would be
visible in our spectra.  This fit could only be improved by increasing
the size (and mass) of the brightest star to a value inconsistent with
the surface gravity inferred from our spectroscopy.

\begin{figure}[p]
\epsscale{1.0}
\plotone{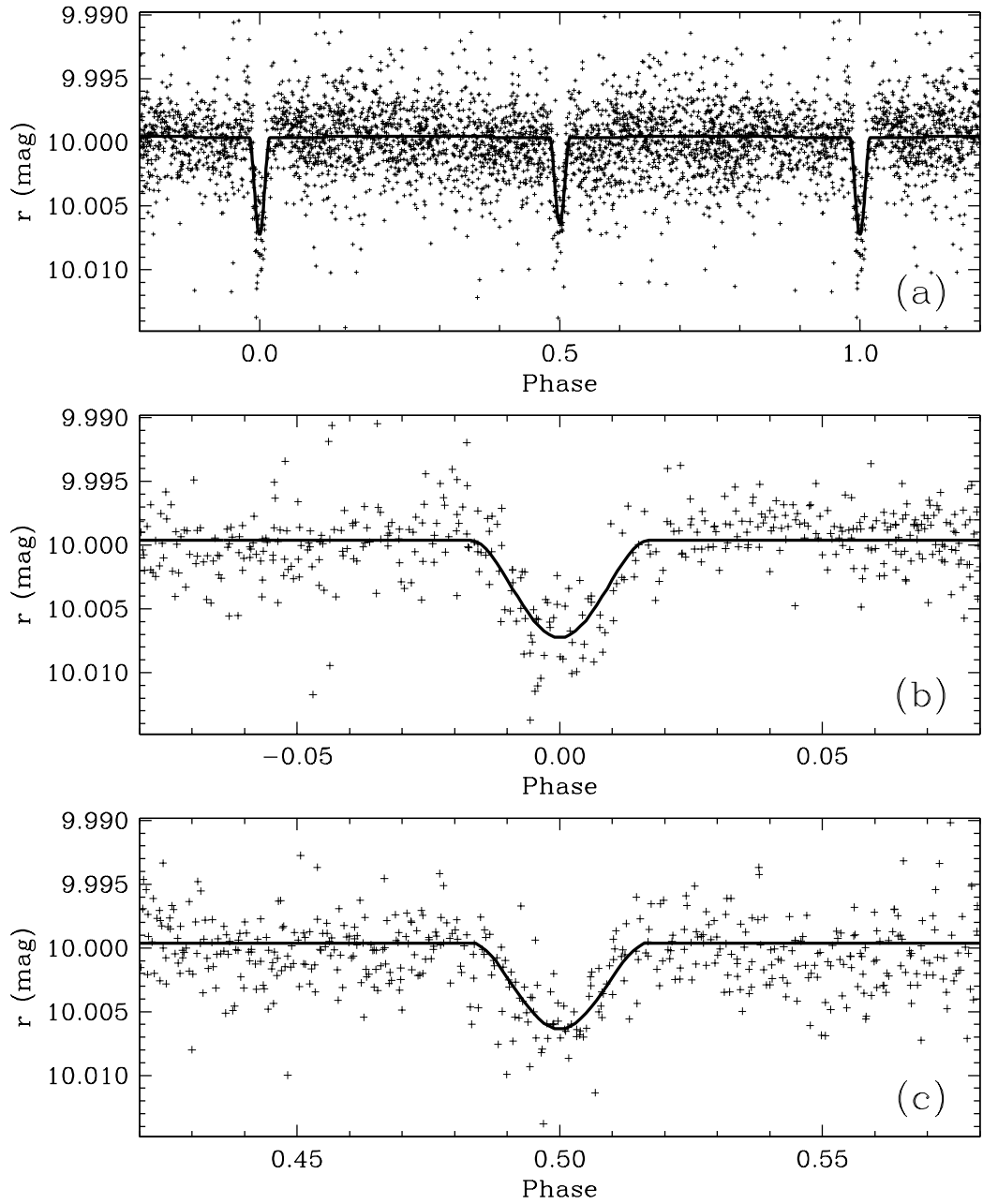}
\caption{TrES $r$--band photometry of \mbox{GSC\,03885--00829}, folded
  using a period twice that of our candidate planet (i.e., $2 \times
  1.441$ days). (a) Best--fit blend model consisting of the bright F
  star and a pair of eclipsing K dwarfs (see text). The theoretical
  curve indicates a small difference in depth between the primary and
  secondary eclipse; (b) Enlargement around the primary eclipse; (c)
  Enlargement around the secondary eclipse. The observed duration of
  the eclipses is well reproduced by the model.}
\label{fig:MMmodel}
\end{figure}

The periods for our transit candidates are calculated assuming each
observed transit is of equal depth. However, when a candidate is in
fact a blended eclipsing binary, the possibility exists that the
primary and secondary eclipses are similar enough in depth to be
confused by our period-finding technique. The BLS algorithm may derive
a best fit period for such a system that is half of the true value.
Therefore, we explored blend scenarios in which the period is $2
\times 1.441$ days. Figure~\ref{fig:MMmodel} shows the result of our
best fit to the TrES $r$-band data, in which the eclipsing binary is
composed of two late K stars with masses of 0.67\,M$_{\sun}$ and
0.64\,M$_{\sun}$, with an orbital inclination of $84\arcdeg$ to the
line of sight.  The model indicates a slight difference in eclipse
depths (see Figure~\ref{fig:MMmodel}a) of about 1 mmag, although this
difference is only marginally visible in observations themselves.
Enlargements of the two eclipse regions are displayed in the lower
panels of Figure~\ref{fig:MMmodel}.  According to this fit, the time
of the center of transit previously derived (see Table~\ref{tab:gsc})
is in fact a time of secondary eclipse; all of the photometric
follow-up observations shown in Figure~\ref{fig:multicolor} were taken
during a secondary eclipse.  The brighter of the K stars (i.e., the
primary of the eclipsing binary) has only $\sim$3\% of the light of
the main F star in the optical, and is below our threshold for
spectroscopic detection.  Figure~\ref{fig:isochrones} shows the
location of the three stars (the filled circles) in the H--R diagram.

According to the blend model described above, the eclipse in the $g$
band (a secondary eclipse) is predicted to be shallower than in the
$r$ band (0.35\% versus 0.7\%), as we indeed observe, although the
measured depth (0.4\%) is slightly deeper than predicted. We attribute
this to shortcomings in the isochrones used for the blend modeling,
which are not specifically designed for low-mass stars. In particular,
missing opacity sources and other physical ingredients may affect the
theoretical luminosities in the optical ($V$ or $g$) bands
\citep{Baraffe_Chabrier_Allard:aa:1998a,
  Delfosse_Forveille_Segransan:aa:2000a,
  Chabrier_Baraffe_Allard:preprint:2005a}, whereas the red and near
infrared magnitudes are presumably more reliable. A sign of this is
seen perhaps in the predicted $V-K$ color for the main F star: the
isochrones give $V-K = 1.26$ \citep[in the Johnson system as defined
by][]{Bessell_Brett:pasp:1988a}, bluer than the typical color of a
dwarf of this temperature, $V-K = 1.40$
\citep[e.g.,][]{Bessell_Brett:pasp:1988a}.  Other stellar evolution
models specifically designed for low-mass stars such as those by
\cite{Baraffe_Chabrier_Allard:aa:1998a} appear to give more realistic
colors. Our F star is predicted to have $V-K = 1.41$ according to
those calculations \citep[after transformation of the isochrone $K$
magnitudes from the CIT to the Johnson system,
following][]{Leggett:apjs:1992}, very close to the empirical
value. Unfortunately the \cite{Baraffe_Chabrier_Allard:aa:1998a}
models are not publicly available for the Sloan bands, so we are
unable to use them in our blend modeling.

\begin{figure}[p]
\epsscale{0.7}
\plotone{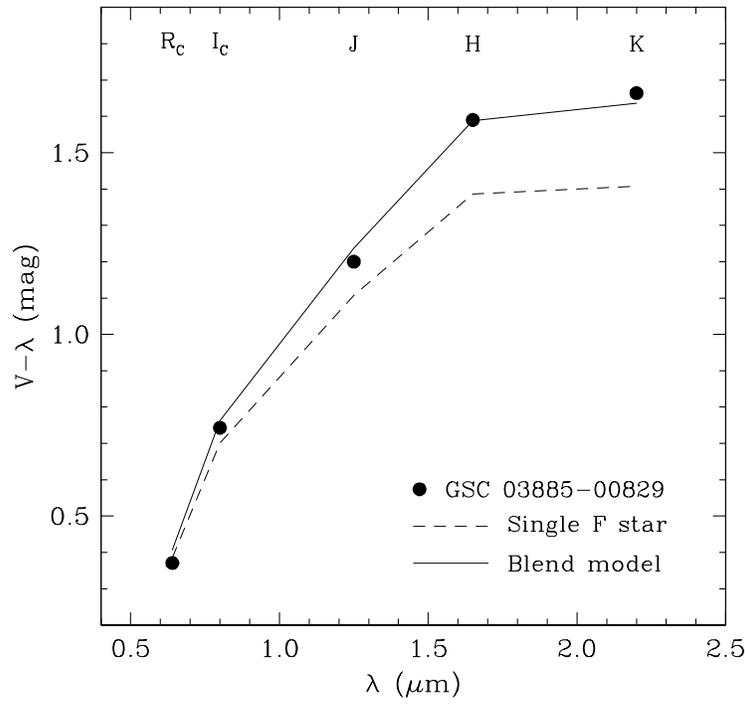}
\caption{Near infrared colors measured for \mbox{GSC\,03885--00829}
  (dots) compared with the theoretical colors of a single F star
  (dashed line) and the colors from our blend model (solid line;
  combined light of three stars). The latter is seen to reproduce the
  observed colors well.}
\label{fig:colors}
\end{figure}

As indicated earlier, infrared magnitudes for \gsc\ are
available from the 2MASS Catalog.  In particular, the measured $V-K$
color of our candidate in the Johnson system (Table~\ref{tab:gsc}) is 
$1.66 \pm 0.02$ 
\citep[using transformations from the 2MASS system
by][]{Carpenter:aj:2001a}. The difference with the color of a single F star
indicates a significant infrared excess of about a quarter of a
magnitude%
\footnote{Hence, even if we underestimated the error in our $V$--band
  photometry (see \S~\ref{sec:photo}) by an order of magnitude, the
  resulting propagation of error would not significantly affect the
  size of this discrepancy.}%
. This in itself can be taken as evidence of contamination
from the light of a later-type object, providing further evidence that
we are dealing with a blend. The computed color of the combined light
of the three stars in our model using the
 \cite{Baraffe_Chabrier_Allard:aa:1998a} isochrones
is $V-K = 1.64$, which agrees quite well with the observations and
supports our interpretation. Other predicted red and infrared colors
also match the measured values reasonably well (see Figure~\ref{fig:colors}).

Although it is quite possible that additional fine--tuning may improve
the small discrepancies noted above in the $g$ band and provide a
near--perfect fit to all observations (to the extent allowed by the
accuracy of the stellar evolution models and observational
uncertainties), our goal here has been to show how subtle the
signatures of a blend can be, and that with careful modeling it is
possible to demonstrate that they are in fact due to a blend scenario,
and therefore to reject the candidate.

\section{Confirmation of Blend Model}
\label{sec:nirspec}

In order to test further our blend hypothesis, we observed
\gsc\ on UT 15 August 2005 with the NIRSPEC
infrared spectrograph at the Keck Observatory.  We observed
the target in the $K$--band spectral region centered near 2.293\,$\mu$m,
to search for the presence of features from the 12CO 2--0 bandhead.
Such features are very weak for mid--G--type stars, and absent for
stars with spectral types earlier than G0.  Hence, the detection
of such features would indicate the presence of a cool stellar
photosphere, as predicted by our blend scenario.

We used a 3--pixel--wide slit, which yields a spectral resolution of
approximately 25,000.  We gathered two 4--minute exposures,
between which we nodded along the slit by roughly $5\arcsec$.
We differenced the two exposures to subtract the sky emission
and any pixel--dependent detector bias.  We extracted the
order spanning the location of the 12CO 2--0 bandhead
by summing over a 15--pixel--wide band centered on the
peak of the instrumental profile.  A small number of
values in the extracted 1--dimensional spectrum were corrupted
due to bad pixels in the infrared detector.  We replaced
these values (30 out of 1024 pixels) by interpolation.
This region contains a large number of telluric methane features.  We
produced a model of these features by modifying the electronic version of
KPNO/FTS telluric spectrum \citep{Livingston_Wallace:NSO:1991a} for airmass,
wavelength--solution, and instrumental point--spread function.
Dividing our extracted spectrum by this model yields the
stellar spectrum corrected for telluric absorption.

\begin{figure}[p]
\epsscale{1.0}
\plotone{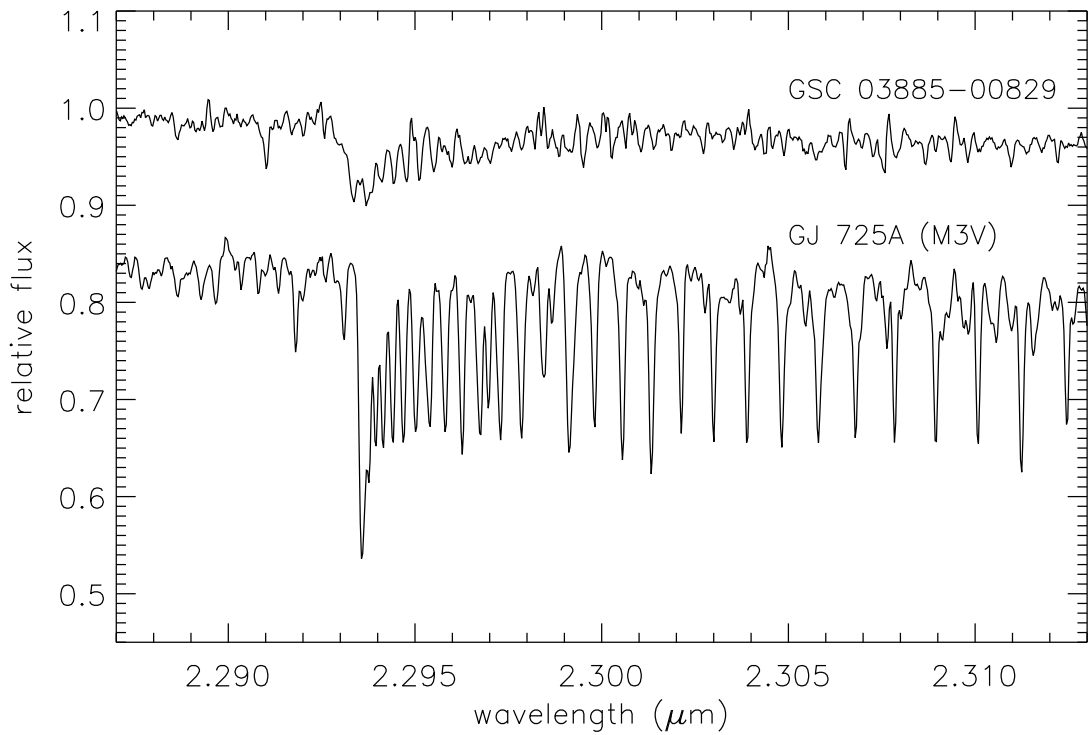}
\caption{$K$--band spectra of \mbox{GSC\,03885--00829} and a nearby
  M3V star (\mbox{GJ\,725A}), obtained with the NIRSPEC spectrograph
  at the Keck Observatory. The M dwarf spectrum, with strong CO
  features, is shown for comparison with our target spectrum, which
  displays the 12CO 2--0 band head at approximately $2.293\,\mu$m. The
  spectrum must therefore include the light of a star with a spectral
  type later than the F dwarf target star, such as the K dwarfs of our
  blend model. The presence of the CO feature thus rules out the
  possibility of a planetary companion.}
\label{fig:nirspec}
\end{figure}

In Figure~\ref{fig:nirspec}, we plot the resulting spectrum, as well as a
spectrum of the nearby M3V star \mbox{GJ\,725A} (which
shows very prominent CO features) for comparison. The relative intensity 
between the individual CO features at 2.3 um does not
differ for these two types of stars, although the overall amplitude of the
features is reduced for hotter temperatures. We did not attempt to match the
spectral type of the NIRSPEC data quantitatively, since our primary goal
is to exclude the planetary hypothesis. The spectrum of 
\gsc\ clearly shows the 12CO 2--0 bandhead near 2.293\,$\mu$m; 
the relative depth of the band head
is approximately 6\%. The detection of this feature confirms
the presence of a cool stellar photosphere. Also, for the late K dwarfs of
our model, the 12CO 2--0 band head has a depth of approximately 30\%,
and the K dwarfs in our blend model contribute 25\% of the $K$--band
flux from the system. Hence the expective depth of this band head in
our spectrum is $\sim7.5$\%, in rough agreement with the observed
relative depth. Thus we interpret this feature as originating in the
photospheres of the K--stars of the binary. 

The heliocentric Julian Day at mid--exposure was HJD 2453597.83591,
which corresponds to an orbital phase of 0.09, at which point the
expected velocity separation between the two K-stars is
$89.5\,\mathrm{km\,s^{-1}}$.  The K--stars are likely tidally--locked
and hence their spectral features will have a $v \sin{i} =
12\,\mathrm{km\,s^{-1}}$. Since this is smaller than the predicted
velocity separation at the time of the exposure, we might expect to
resolve the individual components of the observed 12CO 2--0 feature
from the two K--stars of the binary. And indeed the band head in our
spectrum shows two clear peaks of similar depth, with a velocity
separation similar to the predicted separation of the K dwarfs. A more
careful analysis of this spectrum (using for example the
two-dimensional cross-correlation algorithm TODCOR;
\citealt{Zucker_Mazeh:apj:1994a}) should recover the components more
precisely. However, for the purposes of this paper, it is enough to
identify the presence of the light from an eclipsing K binary system
in our spectrum, which rules out the transiting planet hypothesis.

\section{Discussion}
\label{sec:dis}

The difficult task of eliminating any contamination resulting in a
false transit signal is the primary challenge currently facing wide
field transit surveys. Developing this experience will not only enable
us to make firm detections of transiting Jupiters, but will be
extremely valuable as we search for Earth--sized planets outside our
solar system with NASA's {\it Kepler} mission. As with the
ground--based wide--field surveys, the challenge with Kepler will not
only be obtaining the photometric precision necessary to observe these
minute signals, but rejecting all other possible causes of these
eclipses as well. 

We have presented here one of our disappointments: a candidate that
passed all of our initial photometric and spectroscopic tests, but was
later shown to be a result of contaminated light from an eclipsing
binary. As such, it highlights the difficulty in rejecting all false
positives from a transit survey. As the components of this eclipsing
binary are both faint K dwarfs, the resultant radial velocity
variations in the light blended with that from the nearby F star are
not detectable. Also, the resultant color dependence of the blended
light curve is small, though observable, and the system displays a
color redder than that of an isolated F star.

The TrES survey readily produces on the order of 10 candidate
transiting planets from each selected field of view with
15,000--25,000 stars, the majority of these candidates proving to be
false positives.  Some of these will mimic the expected properties of
a planetary system quite closely. Based on our experience of frequent
blends, where the telltale indicators of the stellar components are
often masked, spectroscopic and even multi--color photometric
follow--up is insufficient to confirm the planetary nature of a
candidate. An attempt must be made to interpret the observations as
those of a blended eclipsing binary, and this interpretation rejected
only if the observations are not in agreement.  Having meticulously
examined the evidence, we can commit to obtain the radial velocity
orbit of this firm candidate through high resolution spectroscopy with
a high signal--to--noise ratio. It should be emphasized that
determining the mass of the candidate from such observations is a
necessary step in identifying a transiting planet.  The methods to
reject astrophysical false positives presented here and by other
authors cannot be used to confirm the planetary nature of a candidate;
rather they increase the yield of planets from the resource--intensive
high--dispersion spectroscopy required for such a confirmation.

\acknowledgments

FTOD and DC thank Lynne Hillenbrand for her advice and continuing
support of this thesis work. We thank the anonymous referee for
several constructive comments. We thank Travis Metcalfe, Nicholas
Nell, Katherine Brown, Courtney Hoskins, and William Kent for their
able assistance during our observations of \gsc\ with the Sommers
Bausch Observatory $24\arcsec$ telescope. We are grateful for the help
of Jose Manuel Almenara and Breezy Oca\~{n}a with the observations
made using the IAC--80 telescope (operated by the Instituto de
Astrof\'\i sica de Canarias at the Observatorio del Teide). We also
thank Russell White for his assistance in gathering our NIRSPEC
observations. The authors wish to recognize and acknowledge the very
significant cultural role and reverence that the summit of Mauna Kea
has always had within the indigenous Hawaiian community.  We are most
fortunate to have the opportunity to conduct observations from this
mountain. This material is based upon work supported by the National
Aeronautics and Space Administration under grant NNG05GJ29G, issued
through the Origins of Solar Systems Program. GT acknowledges partial
support for this work from NASA Origins grant NNG04LG89G. We
acknowledge support for this work from NASA's {\it Kepler}
mission. This research has made use of the SIMBAD database, operated
at CDS, Strasbourg, France, and NASA's Astrophysics Data System
Bibliographic Services.  This publication also utilizes data products
from the Two Micron All Sky Survey, which is a joint project of the
University of Massachusetts and the Infrared Processing and Analysis
Center/California Institute of Technology, funded by the National
Aeronautics and Space Administration and the National Science
Foundation. This research has made use of the
\anchor{http://www.nofs.navy.mil/data/fchpix/}{USNOFS Image and
  Catalogue Archive}%
\footnote{\url{http://www.nofs.navy.mil/data/fchpix/}}%
\ operated by the United States Naval Observatory, Flagstaff Station.

{\it Facilities:} FLWO:1.2m, Hall

\bibliographystyle{apj}
\bibliography{apjmnemonic,mybib.planets}

\begin{thebibliography}{}

\bibitem[\protect\citeauthoryear{{Alard}}{{Alard}}{2000}]{Alard:aas:2000a}
{Alard}, C. 2000, A\&AS, 144, 363

\bibitem[\protect\citeauthoryear{{Alonso} et~al.}{{Alonso}
  et~al.}{2004a}]{Alonso_Brown_Torres:apjl:2004a}
{Alonso}, R., et~al. 2004a, ApJ, 613, L153

\bibitem[\protect\citeauthoryear{{Alonso} et~al.}{{Alonso}
  et~al.}{2004b}]{Alonso_Deeg_Brown:an:2004a}
{Alonso}, R., {Deeg}, H.~J., {Brown}, T.~M.,  \& {Belmonte}, J.~A. 2004b,
  Astron. Nachr., 325, 594

\bibitem[\protect\citeauthoryear{{Bakos} et~al.}{{Bakos}
  et~al.}{2002}]{Bakos_Lazar_Papp:pasp:2002a}
{Bakos}, G.~{\'A}., {L{\'a}z{\'a}r}, J., {Papp}, I., {S{\'a}ri}, P.,  \&
  {Green}, E.~M. 2002, PASP, 114, 974

\bibitem[\protect\citeauthoryear{{Baraffe} et~al.}{{Baraffe}
  et~al.}{1998}]{Baraffe_Chabrier_Allard:aa:1998a}
{Baraffe}, I., {Chabrier}, G., {Allard}, F.,  \& {Hauschildt}, P.~H. 1998,
  A\&A, 337, 403

\bibitem[\protect\citeauthoryear{{Bessell} \& {Brett}}{{Bessell} \&
  {Brett}}{1988}]{Bessell_Brett:pasp:1988a}
{Bessell}, M.~S.,  \& {Brett}, J.~M. 1988, PASP, 100, 1134

\bibitem[\protect\citeauthoryear{{Boss}}{{Boss}}{1997}]{Boss:sc:1997a}
{Boss}, A.~P. 1997, Science, 276, 1836

\bibitem[\protect\citeauthoryear{{Brown}}{{Brown}}{2003}]{Brown:apjl:2003a}
{Brown}, T.~M. 2003, ApJ, 593, L125

\bibitem[\protect\citeauthoryear{{Carpenter}}{{Carpenter}}{2001}]{Carpenter:aj%
:2001a}
{Carpenter}, J.~M. 2001, AJ, 121, 2851

\bibitem[\protect\citeauthoryear{{Chabrier} et~al.}{{Chabrier}
  et~al.}{2005}]{Chabrier_Baraffe_Allard:preprint:2005a}
{Chabrier}, G., {Baraffe}, I., {Allard}, F.,  \& {Hauschildt}, P.~H. 2005, in
  ASP Conf. Ser. : Resolved Stellar Populations, ed. D.~{Valls-Gabaud} \&
  M.~{Chavez}, in press (astro-ph/0509798)

\bibitem[\protect\citeauthoryear{{Charbonneau} et~al.}{{Charbonneau}
  et~al.}{2005}]{Charbonneau_Allen_Megeath:apj:2005a}
{Charbonneau}, D., et~al. 2005, ApJ, 626, 523

\bibitem[\protect\citeauthoryear{{Charbonneau} et~al.}{{Charbonneau}
  et~al.}{2004}]{Charbonneau_Brown_Dunham:AIP:2004a}
{Charbonneau}, D., {Brown}, T.~M., {Dunham}, E.~W., {Latham}, D.~W., {Looper},
  D.~L.,  \& {Mandushev}, G. 2004, in AIP Conf. Proc. 713: The Search for Other
  Worlds, 151

\bibitem[\protect\citeauthoryear{{Charbonneau} et~al.}{{Charbonneau}
  et~al.}{2000}]{Charbonneau_Brown_Latham:apjl:2000a}
{Charbonneau}, D., {Brown}, T.~M., {Latham}, D.~W.,  \& {Mayor}, M. 2000, ApJ,
  529, L45

\bibitem[\protect\citeauthoryear{{Charbonneau} et~al.}{{Charbonneau}
  et~al.}{2002}]{Charbonneau_Brown_Noyes:apj:2002a}
{Charbonneau}, D., {Brown}, T.~M., {Noyes}, R.~W.,  \& {Gilliland}, R.~L. 2002,
  ApJ, 568, 377

\bibitem[\protect\citeauthoryear{{Charbonneau} et~al.}{{Charbonneau}
  et~al.}{2006}]{Charbonneau_Winn_Latham:apj:2006a}
{Charbonneau}, D., et~al. 2006, ApJ, 636, 445

\bibitem[\protect\citeauthoryear{{Creevey} et~al.}{{Creevey}
  et~al.}{2005}]{Creevey_Benedict_Brown:apjl:2005a}
{Creevey}, O.~L., et~al. 2005, ApJ, 625, L127

\bibitem[\protect\citeauthoryear{{Cutri} et~al.}{{Cutri}
  et~al.}{2003}]{Cutri_Skrutskie_van-Dyk:VizieR:2003a}
{Cutri}, R.~M., et~al. 2003, VizieR Online Data Catalog, 2246, 0

\bibitem[\protect\citeauthoryear{{Deeg} \& {Doyle}}{{Deeg} \&
  {Doyle}}{2001}]{Deeg_Doyle:2001a}
{Deeg}, H.~J.,  \& {Doyle}, L.~R. 2001, in Third Workshop on Photometry, ed.
  W.~J. {Borucki} \& L.~E. {Lasher}, 85

\bibitem[\protect\citeauthoryear{{Delfosse} et~al.}{{Delfosse}
  et~al.}{2000}]{Delfosse_Forveille_Segransan:aa:2000a}
{Delfosse}, X., {Forveille}, T., {S{\'e}gransan}, D., {Beuzit}, J.-L., {Udry},
  S., {Perrier}, C.,  \& {Mayor}, M. 2000, A\&A, 364, 217

\bibitem[\protect\citeauthoryear{{Deming} et~al.}{{Deming}
  et~al.}{2005}]{Deming_Seager_Richardson:nat:2005a}
{Deming}, D., {Seager}, S., {Richardson}, L.~J.,  \& {Harrington}, J. 2005,
  Nature, 434, 740

\bibitem[\protect\citeauthoryear{{Dunham} et~al.}{{Dunham}
  et~al.}{2004}]{Dunham_Mandushev_Taylor:pasp:2004a}
{Dunham}, E.~W., {Mandushev}, G.~I., {Taylor}, B.~W.,  \& {Oetiker}, B. 2004,
  PASP, 116, 1072

\bibitem[\protect\citeauthoryear{{Girardi} et~al.}{{Girardi}
  et~al.}{2000}]{Girardi_Bressan_Bertelli:aas:2000a}
{Girardi}, L., {Bressan}, A., {Bertelli}, G.,  \& {Chiosi}, C. 2000, A\&AS,
  141, 371

\bibitem[\protect\citeauthoryear{{Henry} et~al.}{{Henry}
  et~al.}{2000}]{Henry_Marcy_Butler:apj:2000a}
{Henry}, G.~W., {Marcy}, G.~W., {Butler}, R.~P.,  \& {Vogt}, S.~S. 2000, ApJ,
  529, L41

\bibitem[\protect\citeauthoryear{{H{\o}g} et~al.}{{H{\o}g}
  et~al.}{2000}]{Hog_Fabricius_Makarov:aa:2000a}
{H{\o}g}, E., et~al. 2000, A\&A, 355, L27

\bibitem[\protect\citeauthoryear{{Kov{\'a}cs}, {Zucker}, \&
  {Mazeh}}{{Kov{\'a}cs} et~al.}{2002}]{Kovacs_Zucker_Mazeh:aa:2002a}
{Kov{\'a}cs}, G., {Zucker}, S.,  \& {Mazeh}, T. 2002, A\&A, 391, 369

\bibitem[\protect\citeauthoryear{{Kurucz}}{{Kurucz}}{1993}]{Kurucz:ATLAS9:1993%
a}
{Kurucz}, R. 1993, {ATLAS9 Stellar Atmosphere Programs and 2 km/s grid},
  {Kurucz CD--ROM} No.~13 ({Cambridge: SAO})

\bibitem[\protect\citeauthoryear{{Landolt}}{{Landolt}}{1992}]{Landolt:aj:1992a}
{Landolt}, A.~U. 1992, AJ, 104, 340

\bibitem[\protect\citeauthoryear{{Latham}}{{Latham}}{1992}]{Latham:ASP:1992a}
{Latham}, D.~W. 1992, in ASP Conf. Ser. 32: IAU Colloq. 135: Complementary
  Approaches to Double and Multiple Star Research, 110

\bibitem[\protect\citeauthoryear{{Latham}}{{Latham}}{2003}]{Latham:ASP:2003a}
{Latham}, D.~W. 2003, in ASP Conf. Ser. 294: Scientific Frontiers in Research
  on Extrasolar Planets, 409

\bibitem[\protect\citeauthoryear{{Laughlin} et~al.}{{Laughlin}
  et~al.}{2005}]{Laughlin_Wolf_Vanmunster:apj:2005a}
{Laughlin}, G., {Wolf}, A., {Vanmunster}, T., {Bodenheimer}, P., {Fischer}, D.,
  {Marcy}, G., {Butler}, P.,  \& {Vogt}, S. 2005, ApJ, 621, 1072

\bibitem[\protect\citeauthoryear{{Leggett}}{{Leggett}}{1992}]{Leggett:apjs:199%
2}
{Leggett}, S.~K. 1992, ApJS, 82, 351

\bibitem[\protect\citeauthoryear{{Livingston} \& {Wallace}}{{Livingston} \&
  {Wallace}}{1991}]{Livingston_Wallace:NSO:1991a}
{Livingston}, W.,  \& {Wallace}, L. 1991, {An atlas of the solar spectrum in
  the infrared from 1850 to 9000 cm-1 (1.1 to 5.4 micrometer)} ({NSO Technical
  Report 91-001; Tucson: NSO})

\bibitem[\protect\citeauthoryear{{Mandushev} et~al.}{{Mandushev}
  et~al.}{2005}]{Mandushev_Torres_Latham:apj:2005a}
{Mandushev}, G., et~al. 2005, ApJ, 621, 1061

\bibitem[\protect\citeauthoryear{{Monet} et~al.}{{Monet}
  et~al.}{2003}]{Monet_Levine_Canzian:aj:2003a}
{Monet}, D.~G., et~al. 2003, AJ, 125, 984

\bibitem[\protect\citeauthoryear{{Pollack}}{{Pollack}}{1984}]{Pollack:araa:198%
4a}
{Pollack}, J.~B. 1984, ARA\&A, 22, 389

\bibitem[\protect\citeauthoryear{{Pollack} et~al.}{{Pollack}
  et~al.}{1996}]{Pollack_Hubickyj_Bodenheimer:icarus:1996a}
{Pollack}, J.~B., {Hubickyj}, O., {Bodenheimer}, P., {Lissauer}, J.~J.,
  {Podolak}, M.,  \& {Greenzweig}, Y. 1996, Icarus, 124, 62

\bibitem[\protect\citeauthoryear{{Sato} et~al.}{{Sato}
  et~al.}{2005}]{Sato_Fischer_Henry:apj:2005a}
{Sato}, B., et~al. 2005, ApJ, 633, 465

\bibitem[\protect\citeauthoryear{{Stetson}}{{Stetson}}{1987}]{Stetson:pasp:198%
7a}
{Stetson}, P.~B. 1987, PASP, 99, 191

\bibitem[\protect\citeauthoryear{{Stetson}}{{Stetson}}{1992}]{Stetson:ASP:1992%
a}
{Stetson}, P.~B. 1992, in ASP Conf. Ser. 25: Astronomical Data Analysis
  Software and Systems I, ed. D.~M. {Worrall}, C.~{Biemesderfer}, \&
  J.~{Barnes}, 297

\bibitem[\protect\citeauthoryear{{Street} et~al.}{{Street}
  et~al.}{2003}]{Street_Pollaco_Fitzsimmons:ASP:2003a}
{Street}, R.~A., et~al. 2003, in ASP Conf. Ser. 294: Scientific Frontiers in
  Research on Extrasolar Planets, 405

\bibitem[\protect\citeauthoryear{{Struve}}{{Struve}}{1952}]{Struve:obs:1952a}
{Struve}, O. 1952, Observatory, 72, 199

\bibitem[\protect\citeauthoryear{{Tody}}{{Tody}}{1993}]{Tody:1993a}
{Tody}, D. 1993, in ASP Conf. Ser. 52: Astronomical Data Analysis Software and
  Systems II, ed. R.~J. {Hanisch}, R.~J.~V. {Brissenden}, \& J.~{Barnes}, 173

\bibitem[\protect\citeauthoryear{{Torres} et~al.}{{Torres}
  et~al.}{2004a}]{Torres_Konacki_Sasselov:apj:2004a}
{Torres}, G., {Konacki}, M., {Sasselov}, D.~D.,  \& {Jha}, S. 2004a, ApJ, 609,
  1071

\bibitem[\protect\citeauthoryear{{Torres} et~al.}{{Torres}
  et~al.}{2004b}]{Torres_Konacki_Sasselov:apj:2004b}
{Torres}, G., {Konacki}, M., {Sasselov}, D.~D.,  \& {Jha}, S. 2004b, ApJ, 614,
  979

\bibitem[\protect\citeauthoryear{{Vidal-Madjar} et~al.}{{Vidal-Madjar}
  et~al.}{2003}]{Vidal-Madjar_Lecavelier-des-Etangs_Desert:nat:2003a}
{Vidal-Madjar}, A., {Lecavelier des Etangs}, A., {D{\'e}sert}, J.-M.,
  {Ballester}, G.~E., {Ferlet}, R., {H{\'e}brard}, G.,  \& {Mayor}, M. 2003,
  Nature, 422, 143

\bibitem[\protect\citeauthoryear{{Zucker} \& {Mazeh}}{{Zucker} \&
  {Mazeh}}{1994}]{Zucker_Mazeh:apj:1994a}
{Zucker}, S.,  \& {Mazeh}, T. 1994, ApJ, 420, 806

\end{thebibliography}

\begin{deluxetable}{lc}
\tablewidth{0pt}
\tablecaption{Data for \mbox{GSC\,03885--00829} \label{tab:gsc}}
\tablehead{\colhead{Parameter} & \colhead{Value} }
\startdata
R.A. \phm{00.} (J2000)    & \phm{000} \phn \phs $16^{\rm h} 52^{\rm m} 33\fs 7$ \\
Decl. \phm{00} (J2000)  & \phm{000} $+57\arcdeg 58\arcmin 27\arcsec$ \\
GSC & \phm{0000} \phn 03885--00829 \\
2MASS & \phm{0000} 16523368+5758262 \\ 
 & \\
$V$ \tablenotemark{a} \phm{$-R_{\rm C}$} (mag) & \phm{0000} $10.465 \pm 0.001$ \\
$B-V$  \tablenotemark{a} \phd\phd (mag) & \phm{0000} \phn$0.612 \pm 0.001$ \\
$V-R_{\rm C}$  \tablenotemark{a} (mag) & \phm{0000} \phn$0.371 \pm 0.003$ \\
$V-I_{\rm C}$  \tablenotemark{a} \phd (mag) & \phm{0000} \phn$0.743 \pm 0.001$ \\
$J$  \tablenotemark{b} \phm{$-R_{\rm C}.$} (mag) & \phm{0000} \phn$9.197 \pm 0.018$ \\
$J-H$  \tablenotemark{b} \phd \phd (mag) & \phm{0000} \phn$0.337 \pm 0.023$ \\
$J-K_{s}$  \tablenotemark{b} \phd (mag) & \phm{0000} \phn$0.440 \pm 0.026$ \\
 & \\
Period \tablenotemark{c} \phm{000} (d) & \phm{00000} $2.88244 \pm 0.00046$ \\
$T_{2}$ \tablenotemark{d} \phm{0000} (HJD) & $2453529.833 \pm 0.009$ \\
Depth \phm{.} ($r$ mag) & \phm{00000000.} $0.006 \pm  0.003 $  
\enddata
\tablenotetext{a}{See \S\ref{sec:photo} for a discussion of errors.}
\tablenotetext{b}{From the 2MASS Catalog \citep{Cutri_Skrutskie_van-Dyk:VizieR:2003a}.}
\tablenotetext{c}{The period of the suspected candidate planet was $1\fd44122 \pm 0\fd00023$.}
\tablenotetext{d}{The time of the \textit{secondary} eclipse of the binary, and the central transit time of the candidate.}
\end{deluxetable}

\begin{figure}[p]
\epsscale{0.5}
\plotone{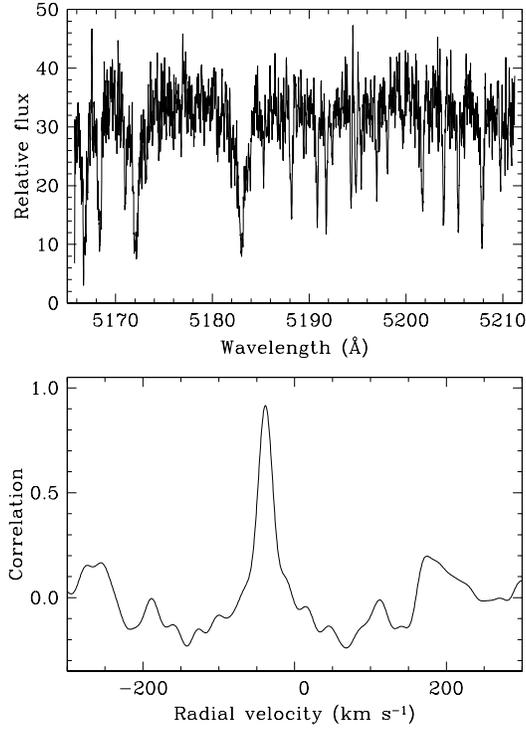}
\caption{Sample spectrum of \mbox{GSC\,03885--00829} (which includes the 
\ion{Mg}{1}~b triplet) and the corresponding cross-correlation function.}
\label{fig:spectrum}
\end{figure}

\end{document}